\documentclass[runningheads]{llncs}
\usepackage[hidelinks]{hyperref}
\usepackage{xurl}
\usepackage[T1]{fontenc}
\usepackage{graphicx}
\usepackage{xcolor}

\newcommand{\todo}[1]{}

\usepackage{nicefrac}
\usepackage{url}
\usepackage{amsmath}
\usepackage{cite}
\usepackage{tikz}

\usepackage{version}
\excludeversion{extended}
\includeversion{conference}

\usepackage{listings}
\usepackage{amssymb}
\definecolor{keywordcolor}{rgb}{0,0,0}
\definecolor{tacticcolor}{rgb}{0,0,0}
\definecolor{commentcolor}{rgb}{0,0,0}
\definecolor{symbolcolor}{rgb}{0,0,0}
\definecolor{sortcolor}{rgb}{0,0,0}
\definecolor{attributecolor}{rgb}{0,0,0}

\def\orcid#1{\smash{\href{http://orcid.org/#1}{\protect\raisebox{-1.25pt}{\protect\includegraphics{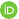}}}}}

\DeclareMathOperator{\trace}{tr}

\spnewtheorem*{example*}{Example}{\bfseries}{}

\begin{document}
\title{Verified reductions for optimization}
\author{Alexander Bentkamp\inst{1,2}\orcid{0000-0002-7158-3595} \and
Ramon Fern\'andez Mir\inst{3}\orcid{0000-0001-7242-5532} \and
Jeremy Avigad\inst{4}\orcid{0000-0003-1275-315X}}
\authorrunning{Bentkamp et al.}
\institute{
Heinrich-Heine-Universität Düsseldorf, Germany 
\and
State Key Laboratory of Computer Science,
Institute of Software,\\
Chinese Academy of Sciences, Beijing, China\\
\and
School of Informatics, University of Edinburgh, Edinburgh, UK\\
\and
Carnegie Mellon University, Pittsburgh, PA, USA}
\maketitle              
\begin{abstract}

Numerical and symbolic methods for optimization are used extensively in engineering, industry, and finance.
Various methods are used to reduce problems of interest to ones that
are amenable to solution by these methods.
We develop a framework for designing and applying such reductions, using the Lean
programming language and interactive proof assistant.
Formal verification makes the process more reliable,
and the availability of an interactive framework and ambient mathematical library
provides a robust environment for constructing the reductions
and reasoning about them.

\end{abstract}
\section{Introduction}
\label{section:introduction}

Optimization problems and constraint satisfaction problems are ubiquitous in
engineering, industry, and finance.
These include the problem of finding an element of $\mathbb{R}^n$ satisfying
a finite set of constraints
or determining that the constraints are unsatisfiable;
the problem of bounding the value of an objective function over a domain
defined by such a set of constraints;
and the problem of finding a value of the domain that maximizes (or minimizes)
the value of an objective function.
Linear programming,
revolutionized by Dantzig's introduction of the simplex algorithm in 1947,
deals with the case in which the constraints and objective function are linear.
The development of interior point methods in the 1980s
allows for the efficient solution of problems defined by convex constraints
and objective functions, which gives rise to the field of
convex programming \cite{boyd-vandenberghe-2014-convex,nesterov-2018-convex-optimization,vishnoi-2021-algorithms}.
Today there are numerous back-end solvers for
convex optimization problems, including MOSEK \cite{mosek},
SeDuMi \cite{sturm-1999-using}, and Gurobi \cite{gurobi}.
They employ a variety of methods,
each with its own particular strengths and weaknesses.
(See \cite[Section 1.2]{agrawal-et-al-2018-cvxpy} for an overview.)

Using such software requires interpreting the problem one wants to solve
in terms of one or more associated optimization problems.
Often, this is straightforward; proving the safety of an engineered system might require
showing that a certain quantity remains within specified bounds,
and an industrial problem might require determining optimal or
near-optimal allocation of certain resources.
Other applications are less immediate.
For example, proving an interesting mathematical theorem may require a lemma
that bounds some quantity of interest (e.g.~\cite{bachoc-vallentin-2008-kissing}).
Once one has formulated the relevant optimization problems,
one has to transform them into ones that the available software can solve,
and one has to ensure that the conditions under
which the software is designed to work correctly have been met.
Mathematical knowledge and
domain-specific expertise are often needed to transform a problem to match
an efficient convex programming paradigm.
A number of modeling packages then provide front ends that
apply further transformations so that the resulting problem
conforms to a back-end solver's input specification
\cite{diamond-boyd-2016-cvxpy,grant-boyd-2014-cvx,lofberg-2004-yalmip,fu-2020-cvxr,udell-et-al-2014-convexjl}.
The transformed problem is sent to the back-end solver
and the solver produces a response,
which then has to be reinterpreted in terms of the
original problem.

Our goal here is to develop ways of using formal methods
to make the passage from an initial mathematical problem to the use of a
back-end solver more efficient and reliable.
Expressing a mathematical problem in a computational proof assistant
provides clarity by endowing claims with a precise semantics, and
having a formal library at hand enables
users to draw on a body of mathematical facts and reasoning procedures.
These make it possible to verify
mathematical claims with respect to the primitives and rules of
a formal axiomatic foundation,
providing strong guarantees as to their correctness.
Complete formalization places a high burden on practitioners
and often imposes a standard that is higher than users want or need,
but verification is not an all-or-nothing affair:
users
should
have the freedom to decide which results they are
willing to trust and which ones ought to be formally verified.

With respect to the use of optimization software, the soundness of the software itself
is one possible concern.
Checking the correctness of a solution
to a satisfaction problem is easy in principle: one simply plugs the result into
the constraints and checks that they hold.
Verifying the correctness of a bounding problem or optimization problem is
often almost as easy, in principle, since the results are often underwritten
by the existence of suitable certificates
that are output by the optimization tools.
In practice, these tasks are made more difficult
by the fact that floating point calculation can introduce
numerical errors that bear on the correctness of the solution.

Here, instead, we focus on the task of manipulating a problem and reducing it to a form
that a back-end solver can handle.
Performing such transformations in a proof assistant
offers strong guarantees that the results are correct and have
the intended meaning,
and it enables users to perform the transformations interactively or partially,
and thus introspect and explore the results of individual transformation steps.
Moreover, in constructing and reasoning about the transformations,
users can take advantage of an ambient mathematical library,
including a database of functions and their properties.

In Section~\ref{section:reduction:to:conic:form}, we describe the
process that CVXPY and other systems use to transform
optimization problems expressed in the \emph{disciplined convex program} (DCP) framework
to conic form problems that can be sent to solvers like MOSEK \cite{mosek}.
In Section~\ref{section:verifying:the:reduction}, we explain how our implementation in the Lean programming language and proof assistant
\cite{de-moura-et-al-2015-lean,de-moura-ullrich-2021-lean4}
augments that algorithm so that it at the same time produces a formal proof
that the resulting reduction is correct.
DCP relies on a library of basic atoms
that serve as building blocks for reductions, and
in Section~\ref{section:adding:atoms}, we explain how our implementation
makes it possible to add new atoms in a verified way.
In Section~\ref{section:user:defined:reductions}, we provide an example of the way that one can further leverage
the power of an interactive theorem prover to justify the reduction of a problem
that lies outside the DCP framework to one that lies within,
using the mathematical library to verify its correctness.
In Section~\ref{section:connecting:lean},
we describe our interface between Lean and an external solver,
which transforms an exact symbolic representation of a problem into
a floating point approximation. Related work is described in
Section~\ref{section:related:work} and conclusions are
presented in Section~\ref{section:conclusions}.

We have implemented these methods in a prototype, CvxLean.\footnote{ \url{https://github.com/verified-optimization/CvxLean}}
We offer more information about the implementation in Section~\ref{section:conclusions}.
A preliminary workshop paper \cite{avigad-bentkamp-2021-workshop} described our initial plans for this project and
the reduction framework presented here in Section~\ref{section:optimization:problems:and:reductions}.

\section{Optimization problems and reductions}
\label{section:optimization:problems:and:reductions}

The general structure of a minimization problem is expressed in Lean 4 as follows:
\begin{lstlisting}
structure Minimization (D R : Type) :=
  (objFun : D → R)
  (constraints : D → Prop)
\end{lstlisting}
Here the data type \lstinline{D} is the \emph{domain} of the problem and
\lstinline{R} is the data type in which the objective function takes
its values. The field \lstinline{objFun} represents the objective function and \lstinline{constraints} is a predicate on \lstinline{D},
which, in Lean, is represented as a function from \lstinline{D} to propositions:
for every value \lstinline{a} of the domain \lstinline{D}, the
proposition \lstinline{constraints a}, which says that the constraints hold of \lstinline{a},
is either true or false.
The domain \lstinline{D} is often $\mathbb{R}^n$ or a space of matrices,
but it can also be something more exotic, like a space of functions.
The data type \lstinline{R} is typically the real numbers, but in full generality it can
be any type that supports an ordering.
A maximization problem is represented as a minimization problem for the negation of
the objective function.

A \emph{feasible point} for the
minimization problem \lstinline{p} is an element \lstinline{point} of \lstinline{D}
satisfying \lstinline{p.constraints}.
Lean's foundational framework allows us to package the data \lstinline{point}
with the condition that it satisfies those constraints:
\begin{lstlisting}
structure FeasPoint {D R : Type} [Preorder R] (p : Minimization D R) :=
  (point : D)
  (feasibility : p.constraints point)
\end{lstlisting}
The curly and square brackets denote parameters that can generally
be inferred automatically.
A \emph{solution} to the minimization problem \lstinline{p} is a feasible point,
denoted \lstinline{point}, such that
for every feasible point \lstinline{y}
the value of the objective function at \lstinline{point} is smaller than
or equal to the value at \lstinline{y}.
\begin{lstlisting}
structure Solution {D R : Type} [Preorder R] (p : Minimization D R) :=
  (point : D)
  (feasibility : p.constraints point)
  (optimality : ∀ y : FeasPoint p, p.objFun point ≤ p.objFun y.point)
\end{lstlisting}
Feasibility and bounding problems can also be expressed in these terms.
If the objective function is constant (e.g.\ the constant zero function),
a solution to the optimization problem is simply a feasible point.
Given a domain, an objective function, and constraints,
the value \lstinline{b} is a strict lower bound on the value of the objective function over the domain
if and only if the feasibility problem obtained by adding the inequality \lstinline{objFun x ≤ b} to the constraints has no solution.

Lean 4 allows us to implement convenient syntax
for defining optimization problems. For example, the following specifies the problem
of maximizing $\sqrt{x - y}$ subject to the constraints $y = 2x - 3$ and $x^2 \le 2$:
\begin{lstlisting}
optimization (x y : ℝ)
  maximize sqrt (x - y)
  subject to
    c1 : y = 2*x - 3
    c2 : x^2 ≤ 2
    c3 : 0 ≤ x - y
\end{lstlisting}
The third condition, \lstinline{c3}, ensures that the objective
function makes sense and is concave on the domain determined by the constraints.
In some frameworks, like CVXPY, this constraint is seen as implicit in the use
of the expression \lstinline{sqrt (x - y)}, but we currently make it explicit in CvxLean.
Problems can also depend on parameters and background conditions.
For example, we can replace \lstinline{c1} above by
\lstinline{y = a*x - 3} for a parameter \lstinline{a}, and we can replace the objective function by \lstinline{b * sqrt (x - y)} with the background assumption \lstinline{0 < b}.

In Section~\ref{section:user:defined:reductions}, we will consider the
covariance estimation for Gaussian variables, which can be expressed
as follows, for a tuple of sample values \lstinline{y}:
\begin{lstlisting}
optimization (R : Matrix (Fin n) (Fin n) ℝ)
  maximize (∏ i, gaussianPdf R (y i))
  subject to
    c_pos_def : R.posDef
\end{lstlisting}
Here \lstinline{Matrix (Fin n) (Fin n) ℝ} is Lean's representation of the
data type of $n \times n$ matrices over the reals, \lstinline{gaussianPdf}
is the Gaussian probability density function defined in
Section~\ref{section:user:defined:reductions}, and the constraint \lstinline{R.posDef}
specifies that \lstinline{R} ranges over positive definite matrices.

If \lstinline{p} and \lstinline{q} are problems, a \emph{reduction} from \lstinline{p} to \lstinline{q} is a function
mapping any solution to \lstinline{q} to a solution to \lstinline{p}.
The existence of such a reduction means that to solve \lstinline{p} it suffices to solve \lstinline{q}.
If \lstinline{p} is a feasibility problem, it means that the feasibility of \lstinline{q} implies the
feasibility of \lstinline{p}, and, conversely,
that the infeasibility of \lstinline{p} implies the infeasibility of \lstinline{q}.
We can now easily describe what we are after:
we are looking for a system that helps a user reduce a problem \lstinline{p} to
a problem \lstinline{q} that can be solved by an external solver.
(For a bounding problem \lstinline{q}, the goal is to show that the constraints
with the negated bound are infeasible by finding a reduction
from an infeasible problem \lstinline{p}.)
At the same time, we wish to verify the correctness of the reduction,
either automatically or with user interaction.
This will ensure that the results from the external solver really address
the problem that the user is interested in solving.

This notion of a reduction is quite general, and is not restricted to any
particular kind of constraint or objective function.
In the sections that follow, we explain how the notion can be applied to convex
programming.

\section{Reduction to conic form}
\label{section:reduction:to:conic:form}

\emph{Disciplined Convex Programming (DCP)} is a framework for writing
constraints and objective functions in such a way that they can automatically
be transformed into problems that can be handled by particular back-end solvers.
It aims to be flexible enough to express optimization problems in a natural way but restrictive enough to ensure that problems can be transformed to meet the requirements of the solvers.
To start with, the framework guarantees that expressions satisfy the
relevant curvature constraints
\cite{grant-et-al-2006-dcp,agrawal-et-al-2018-cvxpy},
assigning a \emph{role} to each expression:
\begin{itemize}
  \item Constant expressions and variables are affine.
  \item An expression $f(\mathsf{expr}_1, \ldots, \mathsf{expr}_n)$ is affine
    if $f$ is an affine function and for each $i$, $\mathsf{expr}_i$ is affine.
  \item An expression $f(\mathsf{expr}_1, \ldots, \mathsf{expr}_n)$ is convex if $f$ is
    convex and for each $i$, one of the following conditions holds:
    \begin{itemize}
      \item $f$ is increasing in its $i$th argument and $\mathsf{expr}_i$ is
        convex.
      \item $f$ is decreasing in its $i$th argument and $\mathsf{expr}_i$ is
        concave.
      \item $\mathsf{expr}_i$ is affine.
    \end{itemize}
  \item The previous statement holds with ``convex'' and ``concave'' switched.
\end{itemize}
An affine expression is both convex and concave.
Some functions $f$ come with side conditions on the range of arguments for which
such curvature properties are valid; e.g.~$f(x) = \sqrt x$ is concave
and increasing on the domain $\{ x \in \mathbb{R} \mid x \ge 0 \}$.

A minimization problem is amenable to the DCP reduction if, following the rules above, its objective function
is convex and the expressions occurring in its constraints are concave or convex,
depending on the type of constraint.
For example, maximizing $\sqrt{x - y}$ requires minimizing $- \sqrt{x - y}$, and
the DCP rules tell us that the latter is a convex function of $x$ and $y$ on the domain
where $x - y \ge 0$, because $x - y$ is affine,
$\sqrt{\cdot}$ is concave and increasing in its argument,
and negation is affine and decreasing in its argument.

CvxLean registers the properties of atomic functions $f(\bar a)$ in a library of \emph{atoms}.
Each such function $f$ is registered with
a formal representation $\mathsf{expr}_f(\bar{a})$ using
expressions, like \lstinline{x * log x} or \lstinline{log (det A)},
that can refer
to arbitrary functions defined in Lean's library.
The atom also registers the relevant properties of $f$.
The curvature of $f$, $\mathsf{curv}_f$,
has one of the values $\mathsf{convex}$, $\mathsf{concave}$, or $\mathsf{affine}$,
and the monotonicity of the function in each of its arguments is tagged as
$\mathsf{increasing}$, $\mathsf{decreasing}$, or $\mathsf{neither}$.
CvxLean also allows the value $\mathsf{auxiliary}$, which indicates an expression
that serves as a fixed parameter in the sense that it is independent of the
variables in the optimization problem.
Atoms can also come with
\emph{background conditions} $\mathsf{bconds}_f(\bar{a})$,
which are independent of the domain variables,
and \emph{variable conditions} $\mathsf{vconds}_f(\bar{a})$, which
constrain the domain on which the properties hold.
Notably, the
atoms also include \emph{proofs} of properties that are needed to
justify the DCP reduction.

By storing additional information with each atom, a DCP
framework can use the compositional representation of expressions to represent
a problem in a form appropriate to a back-end solver.
For example, solvers like MOSEK expect problems to be posed in a certain
\emph{conic form} \cite{mosek}.
To that end, CVXPY stores a \emph{graph implementation}
for each atomic function $f$,
which is a representation of $f$ as the solution to a conic optimization problem.
By definition, the graph implementation of an atomic function $f$ is
an optimization problem in conic form, given by a list of variables $\bar{v}$,
an objective function $\mathsf{obj}_f(\bar{x}, \bar{v})$,
and a list of constraints $\mathsf{constr}_f(\bar{x}, \bar{v})$,
such that the optimal value of the objective under the constraints is
equal to $f(\bar{x})$ for all $\bar{x}$ in the domain of validity.
For example, for any $x \ge 0$, the concave function $\sqrt x$ can be
characterized as the maximum value of the objective function $\mathsf{obj}(x, t) = t$ satisfying the constraint $\mathsf{constr}(x, t)$ given by
$t^2 \le x$.
Once again, a notable feature of CvxLean is that that the atom comes equipped
with a formal proof of this fact.

The idea is that we can reduce a problem to the required form by iteratively
replacing each application of an atomic function by an equivalent
characterization in terms of the graph implementation.
For example, we can replace a subexpression $\sqrt{x - y}$ by a new variable $t$
and add the constraint $t^2 \le x - y$, provided that the form of the resulting
problem ensures that, for any \emph{optimal} solution to the constraints,
$t$ will actually be \emph{equal} to $\sqrt{x - y}$.
Given a well-formed DCP minimization problem, CvxLean must perform the reduction
and construct a formal proof of the associated claims.
In this section we describe the reduction, and in the next section we
describe the proofs.
\begin{extended}
A more formal description of both are given in the appendix.
\end{extended}
\begin{conference}
A more formal description of both
are given in an extended version of this paper \cite{this:paper:extended}.
\end{conference}

Let $e$ be a well-formed DCP expression.
CvxLean associates to each such expression a tree $T$ whose leaves are
expressions that are affine with respect to the
variables of the optimization problem.
For example, this is the tree associated with the expression
\lstinline{-sqrt (x - y)}:
\begin{center}
\begin{tikzpicture}[yscale=0.8]
  \node {neg (affine, in the role of convex)}
    child {node {sqrt (concave)}
      child {
        node {sub (affine, in the role of concave)} [sibling distance = 4cm]
        child {
          node {\lstinline{x}}
          edge from parent node [left] {increasing \hspace*{0.5em}}
        }
        child {
          node {\lstinline{y}}
          edge from parent node [right] {\hspace*{0.5em} decreasing}
        }
        edge from parent node [right] {increasing}
      }
      edge from parent node [right] {decreasing}
    };
\end{tikzpicture}
\end{center}
Alternatively, we could use a single leaf for
\lstinline{x - y}.
Denoting the variables of the optimization problem by $\bar y$,
we can recursively assign to each node $n$ a subexpression
$\mathsf{oexpr}_n(\bar{y})$ of $e$ that corresponds to the subtree with root $n$.
In the example above, the subexpressions are
\lstinline{x}, \lstinline{y},
\lstinline{x - y}, \lstinline{sqrt (x - y)}, and \lstinline{-sqrt (x - y)}.
To each internal node, we assign a curvature, $\mathsf{convex}$, $\mathsf{concave}$,
or $\mathsf{affine}$, subject to the rules of DCP. An expression that is affine can be
viewed as either convex or concave.
Equalities and inequalities are also atoms; for example, $e_1 \le e_2$
describes a convex set
if and only if $e_1$ is convex and $e_2$ is concave.
A formalization of the DCP rules allows us
to recursively construct formal proofs of these curvature claims,
modulo the conditions and assumptions of the problem. We elaborate on this process
in the next section.

Now consider a well-formed DCP minimization problem with objective function
$o$ and constraints $c_1, \dots, c_n$.
We call these expressions the \emph{components} of the problem.
Recall the following example from the previous section, recast as a minimization problem:
\begin{lstlisting}
optimization (x y : ℝ)
  minimize -sqrt (x - y)
  subject to
    c1 : y = 2*x - 3
    c2 : x^2 ≤ 2
    c3 : 0 ≤ x - y
\end{lstlisting}
Here the components are \lstinline{-sqrt (x - y)}, \lstinline{y = 2*x - 3}, \lstinline{x^2 ≤ 2}, and
\lstinline{0 ≤ x - y}.

First, we assign to each component $c$ an atom tree $T_c$ as described above.
If $\bar{y}$ are the variables of the original problem,
the variables of the reduced problem are $\bar{y} \cup \bar{z}$,
where $\bar{z}$ is a collection of variables consisting of a fresh
set of variables for the graph implementation at each internal node of each tree,
for those atoms whose graph implementations introduce new variables.
To each node $n$ of each atom tree, we assign an expression
$\mathsf{rexpr}_n(\bar{y}, \bar{z})$ in the language of the reduced problem
representing the expression $\mathsf{oexpr}_n(\bar{y})$ in the original problem.
At the leaves, $\mathsf{rexpr}_n(\bar{y}, \bar{z})$ is the same as
$\mathsf{oexpr}_n(\bar{y})$. At internal nodes we use the objective function
of the corresponding atom's graph implementation, applied to the interpretation of the arguments.
The objective of the reduced problem is the expression
assigned to the root of $T_o$.

As far as the constraints of the reduced problem, recall that each internal
node of the original problem corresponds to an atom, which has
a graph implementation. The graph implementation, in turn, is given by
a list of variables $\bar{v}$,
an objective function $\mathsf{obj}_f(\bar{a}, \bar{v})$,
and a list of constraints $\mathsf{constr}_f(\bar{a}, \bar{v})$.
These constraints, applied to the expressions representing the arguments,
are part of the reduced problem.
Moreover, the constraints of the original problem, expressed in terms of the reduced problem,
are also constraints of the reduced problem, with one exception.
Recall that atoms can impose conditions $\mathsf{vconds}_f(\bar{a})$, which
are assumed to be among the constraints of the original
problem and to be \emph{implied} by the graph implementation.
For example, the condition \lstinline{0 ≤ x} is required to characterize $\sqrt{x}$ as the
maximum value of a value $t$ satisfying \lstinline{t^2 ≤ x}, but, conversely, the existence of a \lstinline{t}
satisfying \lstinline{t^2 ≤ x} implies \lstinline{0 ≤ x}.
So a constraint \lstinline{0 ≤ x} that is present in the original problem
to justify the use of \lstinline{sqrt x} can be dropped from the reduced problem.

In the example above, there is a tree corresponding to each of the components
\lstinline{-sqrt (x - y)}, \lstinline{x^2 ≤ 2},
\lstinline{0 ≤ x - y}, and \lstinline{y = 2*x - 3}.
As $n$ ranges over the nodes of these trees, $\mathsf{oexpr}_n(x, y)$
ranges over all the subexpressions of these components, namely,
\lstinline{x}, \lstinline{y}, \lstinline{x - y},
\lstinline{sqrt (x - y)}, \lstinline{-sqrt (x - y)},
\lstinline{x^2}, \lstinline{2}, \lstinline{x^2 ≤ 2},
and so on.
The only atoms whose
graph implementations introduce extra variables are the square root and the square.
Thus, CvxLean introduces the
variable \lstinline{t.0}, corresponding to the expression
\lstinline{sqrt (x - y)},
and the variable \lstinline{t.1}, corresponding to the expression
\lstinline{x^2}.
The values of $\mathsf{rexpr}_n(x, y, t_0, t_1)$ corresponding to some of the expressions
above are as follows:
\begin{center}
\begin{tabular}{c ||c|c|c|c}
$\;\mathsf{oexpr}_n(x, y)\;$ & {\verb~ x - y ~} & {\verb~ sqrt (x - y) ~} & {\verb~ -sqrt (x - y) ~} &
{\verb~ x^2 ~} \\
\hline
$\;\mathsf{rexpr}_n(x, y, t_0, t_1)\;$ & {\verb~ x - y ~} &
{\verb~ t.0 ~} & {\verb~ -t.0 ~}
& {\verb~ t.1 ~}
\end{tabular}
\end{center}
The constraints \lstinline{c1} and \lstinline{c2} of the original problem
translate to cone
constraints \lstinline{c1'} and \lstinline{c2'} on the new variables,
the constraint \lstinline{c3} is implied by the
graph representation of \lstinline{x^2}, and the graph representations of
\lstinline{sqrt (x - y)} and \lstinline{x^2} become new cone constraints
\lstinline{c4'} and \lstinline{c5'}. Thus the reduced problem is as follows:
\begin{lstlisting}
optimization (x y t.0 t.1 : ℝ)
  maximize t.0
  subject to
    c1' : zeroCone (2*x - 3 - y)             -- 2*x - 3 - y = 0
    c2' : posOrthCone (2 - t.1)              -- 2 - t.1 ≥ 0
    c4' : rotatedSoCone 0.5 (x - y) ![t.0]   -- x - y ≥ t.0^2
    c5' : rotatedSoCone t.1 0.5 ![x]         -- t.1 ≥ x^2
\end{lstlisting}
Here, \lstinline{![t.0]} and \lstinline{![x]} denote singleton vectors and the meaning of the cone constraints is annotated in the comments.
For a description of the relevant conic forms, see the MOSEK modeling cookbook
\cite{mosek:cookbook}.

\section{Verifying the reduction}
\label{section:verifying:the:reduction}

The reduction described in the previous section is essentially the same as the
one carried out by CVXPY.
The novelty of CvxLean is that it provides a formal justification that the
reduction is correct.
The goal of this section is to explain how we manage to construct a formal proof
of that claim.
In fact, given a problem $P$ with an objective function $f$, CvxLean constructs a new
problem $Q$ with an objective $g$, together with the following additional pieces of data:
\begin{itemize}
\item a function $\varphi$ from the domain of $P$ to the domain of $Q$ such that
  for any feasible point $x$ of $P$, $\varphi(x)$ is a feasible point of $Q$
  with $g(\varphi(x)) \le f(x)$
\item a function $\psi$ from the domain of $Q$ to the domain of $P$ such that for any
 feasible point $y$ of $Q$, $\psi(y)$ is a feasible point of $P$
 with $f(\psi(y)) \le g(y)$.
\end{itemize}
These conditions guarantee that if $y$ is a solution to $Q$
then $\psi(y)$ is a solution to $P$,
because for any feasible point $x$ of $P$ we have
\[
  f(\psi(y)) \le g(y) \le g(\varphi(x)) \le f(x).
\]
This shows that $\psi$ is a reduction of $P$ to $Q$,
and the argument with $P$ and $Q$ swapped shows that $\varphi$ is a reduction of $Q$ to $P$.
Moreover, whenever $y$ is a solution to $Q$, instantiating $x$ to $\psi(y)$ in the chain
of inequalities implies $f(\psi(y)) = g(y)$.
Similarly, when $x$ is a solution to $P$, we have $g(\varphi(x)) = f(x)$.
So the conditions above imply that $P$ has a solution if and only if $Q$ has a solution,
and when they do, the minimum values of the objective functions coincide.
Below, we will refer to the data $(\varphi, \psi)$ as a \emph{strong equivalence} between the two problems.

To construct and verify such a strong equivalence between the original problem
and the result of applying the transformation described in Section~\ref{section:reduction:to:conic:form},
we need to store additional information with each atom.
Specifically, for each atomic function $f(\bar a)$, that atom must provide solutions
$\mathsf{sol}_f(\bar{a})$ to the graph implementation variables $\bar v$,
as well as formal proofs of the following facts:
\begin{itemize}
  \item The function $f(\bar a)$ satisfies the graph implementation: for each $\bar a$ satisfying the
  conditions $\mathsf{vconds}_f(\bar{a})$, we have:
  \begin{itemize}
    \item \emph{solution feasibility:} $\mathsf{sol}_f(\bar{a})$ satisfies the constraints
  $\mathsf{constr}_f(\bar{a}, \mathsf{sol}_f(\bar{a}))$
    \item \emph{solution correctness:} we have $\mathsf{obj}_f(\bar{a}, \mathsf{sol}_f(\bar{a})) = \mathsf{expr}_f(\bar{a})$ ,
    where $\mathsf{expr}_f(\bar{a})$ is the expression representing $f$.
  \end{itemize}
  \item The function $f(\bar a)$ is the \emph{optimal} solution to the graph implementation, in the following sense.
  Write $\bar{a}' \mathrel{\triangle} \bar{a}$ to express the assumptions that $a'_i \geq a_i$ for increasing arguments to $f$,
  $a'_i \le a_i$ for decreasing arguments, and $a'_i$ and $a_i$ are syntactically
  identical for other arguments. If $f$ is convex and $\bar{a} \mathrel{\triangle} \bar {a}'$,
  we require $\mathsf{obj}_f(\bar{a},\bar{v}) \geq \mathsf{expr}_f(\bar{a}')$ for any
  $\bar{v}$ such that $\mathsf{constr}_f(\bar{a}, \bar{v})$ holds.
  If $f$ is concave and $\bar{a}' \mathrel{\triangle} \bar {a}$,
  we require $\mathsf{obj}_f(\bar{a},\bar{v}) \leq \mathsf{expr}_f(\bar{a}')$ for any
  $\bar{v}$ such that $\mathsf{constr}_f(\bar{a}, \bar{v})$ holds.
  For affine atoms, we require both.
\end{itemize}
Finally, as noted in the previous section, the graph implementation implies the conditions
needed for the reduction. Under the assumptions on $\bar{a}$ and $\bar{a}'$ in the second
case above, we also require a proof of $\mathsf{vconds}_f(\bar{a}')$.
We refer to this as \emph{condition elimination}.

For a concrete example, consider the atom for the concave function $\sqrt{a}$. In that case,
$\mathsf{vconds}(a)$ is the requirement $a \ge 0$, and $\mathsf{expr}(a)$, the Lean representation of the function,
is given by Lean's \lstinline{sqrt} function. The graph implementation adds a new variable $v$.
The only constraint $\mathsf{constr}(a, v)$ is $v^2 \le a$, and the objective function is
$\mathsf{obj}(a, v) = v$. The solution function $\mathsf{sol}(a)$ returns $\sqrt a$ when
$a$ is nonnegative and an arbitrary value otherwise.
The atom for $\sqrt{\cdot}$ stores Lean proofs of all of the following:
\begin{itemize}
  \item solution feasibility: \lstinline{∀ a, 0 ≤ a → (sqrt a)^2 ≤ a}
  \item solution correctness: \lstinline{∀ a, 0 ≤ a → sqrt a = sqrt a}
  \item optimality: \lstinline{∀ v a a', a ≤ a' → v^2 ≤ a → v ≤ sqrt a'}
  \item condition elimination: \lstinline{∀ v a a', a ≤ a' → v^2 ≤ a → 0 ≤ a'}.
\end{itemize}
More precisely, the atom stores the representation of the graph of the
square root function as a cone constraint, and the properties above are expressed
in those terms.
These properties entail that \lstinline{sqrt} is concave,
but we do not need to prove concavity explicitly.

Let the variables $\bar y$ range over the domain of the original problem, $P$,
and let the variables $\bar y, \bar z$ be the augmented list of variables
in the reduced problem, $Q$.
We wish to construct a strong equivalence between $P$ and $Q$.
To that end, we need to define a forward
map $\varphi$ and a reverse map $\psi$.
The definition of $\psi$ is easy: we simply project each tuple
$\bar y, \bar z$ to $\bar y$.
The definition of the forward map, $\varphi$, is more involved,
since we have to map each tuple $\bar y$ of values to an expanded tuple $\bar y, \bar z$.
The values of $\bar y$ remain unchanged, so the challenge is to define, for each
new variable $z$, an expression $\mathsf{interp}_z(\bar y)$ to interpret it.

Recall that for each subexpression $\mathsf{oexpr}_n(\bar y)$ in the original problem,
corresponding to a node $n$,
there is an expression $\mathsf{rexpr}_n(\bar y, \bar w)$ involving new variables
from the reduced problem.
Suppose a node $n$ corresponds to an expression $f(u_1, \ldots, u_n)$ in
the original problem, and the graph implementation of $f$ introduces new
variables $\bar v$.
For each $v_j$, we need to devise an interpretation $\mathsf{interp}_{v_j}(\bar y)$.
To start with, $\mathsf{sol}_f$ provides a solution to $v_j$
in terms of the arguments $u_1, \ldots, u_n$.
For each of these arguments, $\mathsf{rexpr}$ provides a representation in terms of the
variables $\bar y$ and other new variables.
Composing these, we get an expression $\mathsf{e}(\bar y, w_1, \ldots, w_{\ell})$ for $v_j$ in terms of
the variables $\bar y$ of the original problem and new variables $w_1, \ldots, w_{\ell}$. Recursively, we find interpretations $\mathsf{interp}_{w_k}(\bar y)$ of each $w_k$,
and define $\mathsf{interp}_{v_j}(\bar y)$ to be
$\mathsf{e}(\bar{y}, \mathsf{interp}_{w_1}(\bar y), \ldots, \mathsf{interp}_{w_{\ell}}(\bar y))$.
In other words, we read off the interpretation of each new variable of the reduced problem
from the intended solution to the graph equation, which may, in turn, require the
interpretation of other new variables that were previously introduced.

In the end, the forward map $\varphi$ is the function
that maps the variables $\bar y$ in the original problem to the tuple $(\bar y, \mathsf{interp}_{z_1}(\bar y), \dots, \mathsf{interp}_{z_m}(\bar y))$,
where $z_1, \ldots, z_m$ are the new variables.
To show that $(\varphi, \psi)$ is a strong equivalence, we must show that
for any feasible point $\bar y$ of the original problem,
$\varphi(\bar y)$ is a feasible point of the reduced problem.
This follows from the solution correctness requirement above.
We also need to show that if $f(\bar y)$ is the objective function of the
original problem and $g(\bar y, \bar z)$ is the objective function of the reduced problem,
$g(\varphi(\bar y)) \leq f(\bar y)$.
In fact, the solution correctness requirement enables us to prove the stronger
property $g(\varphi(\bar y)) = f(\bar y)$.
Finally, we need to show that for any feasible point $\bar y, \bar z$
of the reduced problem, the tuple $\bar y$ is a feasible point
of the original problem and $f(\bar y) \leq g(\bar y, \bar z)$.
To do that, we recursively use the optimality requirement
to show $\mathsf{rexpr}_{n}(\bar{y},\bar{z})
\geq \mathsf{oexpr}_{n}(\bar{y})$ whenever the node $n$ marks a convex
expression or an affine expression in the role of a convex expression,
and $\mathsf{rexpr}_{n}(\bar{y},\bar{z})
\leq \mathsf{oexpr}_{n}(\bar{y})$ whenever the node $n$ marks a concave
expression or an affine expression in the role of a concave expression.

\begin{extended}
A proof that the maps $\varphi$ and $\psi$ constructed above form a strong
equivalence can be found in the appendix,
\end{extended}
\begin{conference}
  A proof that the maps $\varphi$ and $\psi$ constructed above form a strong
  equivalence can be found in the extended version of this paper \cite{this:paper:extended},
\end{conference}
but it is helpful to work through the example
from Section~\ref{section:reduction:to:conic:form} to get a sense of what
the proof means.
For this example, the forward map is $\varphi(x, y) = (x, y, \sqrt{x - y}, x^2)$ and
the reverse map is $\psi(x, y, t_0, t_1) = (x, y)$.
Assuming that $(x, y)$ is a solution to the original problem,
the fact that $\varphi(x, y)$ satisfies \lstinline{c1'} follows from \lstinline{c1},
the fact that it satisfies \lstinline{c2'} follows from \lstinline{c2},
the fact that it satisfies \lstinline{c4'} and \lstinline{c5'}
follows from the fact that $\sqrt{x - y}$ and $x^2$ are correct
solutions to the graph constraints. In this direction, $g(\varphi(x, y)) = -\sqrt{x - y} = f(x, y)$.
In the other direction, assuming that $(x, y, t_0, t_1)$ is a solution to the
reduced problem, the fact that $(x, y)$ satisfies \lstinline{c1}
follows from \lstinline{c1'}, that fact that it satisfies \lstinline{c2} follows
from \lstinline{c2'} and \lstinline{c5'}, and the fact that is satisfies
\lstinline{c3} follows from \lstinline{c4'}.
Here we have $f(\psi(x, y, t_0, t_1)) = -\sqrt{x - y}$ and $g(x, y, t_0, t_1) = -t_0$, and the fact that the former is less than or equal to the latter follows from \lstinline{c4'}.

\section{Adding atoms}
\label{section:adding:atoms}

One important advantage to using an interactive theorem prover as a basis
for solving optimization problems is that it is possible to extend
the atom library in a verified way.
In a system like CVXPY, one declares a new atom with its graph implementation
on the basis of one's background knowledge or a pen-and-paper proof that
the graph implementation is correct and that the function described has the
relevant properties over the specified domain.
In CvxLean, we have implemented syntax with which any user can
declare a new atom in Lean and provide formal proofs of these facts.
The declaration can be made in any Lean file, and it becomes
available in any file that imports that one as a dependency.
Lean has a build system and package manager that handles dependencies on
external repositories,
allowing a community of users to share such mathematical and computational content.

For example, the declaration of the atom for the logarithm looks as follows:
\begin{lstlisting}
declare_atom log [concave] (x : ℝ)+ : log x :=
  conditions (cond : 0 < x)
  implementationVars (t : ℝ)
  implementationObjective t
  implementationConstraints (c_exp : expCone t 1 x)
  solution (t := log x)
  solutionEqualsAtom by ...
  feasibility (c_exp : by ...)
  optimality by ...
  conditionElimination (cond : by ...)
\end{lstlisting}
The ellipses indicate places that are filled by formal proofs.
Proof assistants like Lean allow users to write such proofs
interactively in an environment that displays proof obligations,
the local context, and error messages, all while the user types.
For example, placing the cursor at the beginning of the
optimality block displays the following goal:
\begin{lstlisting}
x t : ℝ
c_exp : expCone t 1 x
⊢ ∀ (y : ℝ), x ≤ y → t ≤ log y
\end{lstlisting}
In other words, given real values $x$ and $t$ and the relevant
constraint in terms of the exponential cone, we need to prove
that for every $y \ge x$, we have $t \le \log(y)$.

For the example we present in the next section, we had to implement the
\emph{log-determinant} atom \cite[Example 9.5]{boyd-vandenberghe-2014-convex},
whose arguments consist of a natural number $n$ and a matrix
$A \in \mathbb{R}^{n \times n}$.
This function is represented in Lean by the atom expression $\mathsf{expr}_{\mathit{log\text{-}det}} = \text{\lstinline{log (det A)}}$, where the parameter
$n$ is implicit in the type of \lstinline{A}.
The curvature is specified to be concave, the monotonicity in $n$ is {\sf auxiliary}
because we do not support the occurrence of optimization variables in this argument,
and the monotonicity in $A$ is {\sf neither} because the value of $\log(\det A)$ is
neither guaranteed to increase nor guaranteed to decrease as $A$ increases.
(The relevant order here on matrices is elementwise comparison.)
The correctness of the reduction requires the assumption that
$A$ is positive definite.
Following CVXPY, we used the following graph implementation:
  \begin{align*}
      &\text{maximize}&&\sum_i t_i\\
      &\text{over}&& t \in \mathbb{R}^n,\ Y \in \mathbb{R}^{n \times n}\\
      &\text{subject to}&&
        (t, 1, y) \in \mathrm{expcone}\\
      &&&
      \left( \begin{array}{rrrrrr}
          D\phantom{^T} & Z\\
          Z^T & A
          \end{array}\right) \text{ positive semidefinite}
  \end{align*}
Here $y$ is the diagonal of $Y$;
$Z$ is obtained from $Y$ by setting all entries below the diagonal to $0$;
and $D$ is obtained from Y by setting all entries off the diagonal to $0$.
Here, saying that the tuple $(t, 1, y)$ is in the exponential cone means
that $e^{y_i} \ge t_i$ for each $i$.
Our implementation in CvxLean required proving that
this graph implementation is correct.
To do so, we formalized an argument in the MOSEK documentation.\footnote{\url{https://docs.mosek.com/modeling-cookbook/sdo.html\#log-determinant}}
This, in turn, required proving
properties of the Schur complement, triangular matrices,
Gram-Schmidt orthogonalization, and LDL factorization.
Moreover, the argument uses the subadditivity of the determinant function,
for which we followed an argument by Andreas Thom on MathOverflow.\footnote{\url{https://mathoverflow.net/questions/65424/determinant-of-sum-of-positive-definite-matrices/65430\#65430}}

\section{User-defined reductions}
\label{section:user:defined:reductions}

An even more important advantage of using an interactive proof assistant
as a framework for convex optimization is that, with enough work,
users can carry out \emph{any} reduction that can be expressed and justified
in precise mathematical terms. As a simple example,
DCP cannot handle an expression of the form $\mathit{exp}(x) \mathit{exp}(y)$
in a problem, requiring us instead to write it as $\mathit{exp}(x + y)$.
But in CvxLean, we have the freedom to express the problem in the first form
if we prefer to and then verify that the trivial reduction is justified:
\begin{lstlisting}
reduction red/prob :
  optimization (x y : ℝ)
    maximize x + y
    subject to
      h : (exp x) * (exp y) ≤ 10 := by
  conv_constr => rw [←Real.exp_add]
\end{lstlisting}
Here the expression \lstinline{rw [←Real.exp_add]} supplies the short
formal proof that $\mathit{exp}(x + y)$ can be replaced by $\mathit{exp}(x) \cdot \mathit{exp}(y)$.

Of course, this functionality becomes more important as the reductions
become more involved. As a more substantial example, we have implemented
a reduction needed to solve the the covariance estimation problem for
Gaussian variables \cite[pp.\ 355]{boyd-vandenberghe-2014-convex}.
In this problem, we are given $N$ samples $y_1, \dots y_N \in \mathbb{R}^n$
drawn from a Gaussian distribution with zero mean
and unknown covariance matrix $R$.
We assume that the Gaussian distribution is nondegenerate,
so $R$ is positive definite and the distribution has density function
\[
  p_R(y) = (2\pi)^{-\nicefrac{n}{2}}\det(R)^{-\nicefrac{1}{2}}\exp(-y^T R^{-1} y/2).
\]
We want to estimate the covariance matrix $R$
using maximum likelihood estimation, i.e.,
we want to find the covariance matrix that maximizes
the likelihood of observing $y_1, \dots y_N $.
The maximum likelihood estimate for $R$ is the solution to the following problem:
\begin{align*}
  \text{maximize}\;\;\prod_{k=1}^N p_R(y_k)\;\;
  \text{over}\;\; R\;\;
  \text{subject to}\;\; R \text{ positive definite}.
\end{align*}
As stated, this problem has a simple analytic solution, namely, the sample
covariance of $y_1, \ldots, y_n$, but the problem becomes more interesting when
one adds additional constraints, for example, upper and lower matrix bounds on $R$,
or constraints on the condition number of $R$ (see \cite{boyd-vandenberghe-2014-convex}).
We can easily reduce the problem to maximizing the logarithm of the objective
function above,
but that is not a concave function of $R$.
It is, however, a concave function of $S = R^{-1}$, and common constraints on
$R$ translate to convex constraints on $S$.
We can therefore reduce the problem above to the following:
\begin{align*}
  \text{maximize}\;\; \log(  \det(S) ) - \sum_{k=1}^N y_k^T S y_k\;\;
  \text{over}\;\; S \;\;
  \text{subject to}\;\; S \text{ positive definite},
\end{align*}
possibly with additional constraints on $S$.
We express the sum using
the sample covariance $Y = \frac{1}{N}\sum_{k=1}^N y_k y_k^T$ and the trace operator:
\begin{align*}
  &\text{maximize}\;\; \log(  \det(S) ) - N \cdot \trace(YS^T)\;\;
  \text{over}\;\; S\;\;\\
  &\text{subject to}\;\; S \text{ positive definite}
\end{align*}
The problem can then be solved using disciplined convex programming.
The constraint that $S$ is positive definite is eliminated
while applying the graph implementation of $ \log(  \det(S) )$.
We have formalized these facts in Lean and used them to justify the reduction.
An example with an additional sparsity constraints on $R$ can be found in {\tt CvxLean/Examples} in our repository.

\section{Connecting Lean to a conic optimization solver}
\label{section:connecting:lean}

Once a problem has been reduced to conic form, it can be sent to an external back-end solver.
At this point, we must pass from the realm of precise symbolic representations and
formal mathematical objects to the realm of numeric computation with floating point
representations.
We traverse our symbolic expressions, replacing functions on the reals from Lean's
mathematical library with corresponding numeric functions on floats,
for example associating the floating point exponential function \lstinline{Float.exp}
to the real exponential function \lstinline{Real.exp}.
Our implementation makes it easy to declare such associations with the
following syntax:
\lstinline{addRealToFloat : Real.exp := Float.exp}.

This is one area where more verification is possible.
We could use verified libraries for floating point arithmetic \cite{akbarpour-2010-ieee,boldo-filliatre-2007-floating-point,goodloe-et-al-2013-numerical-programs,yu-2013-floating-point}, we could use
dual certificates to verify the results of the external solver,
and we could carry out formal sensitivity analysis to manage and bound errors.
Our current implementation is only designed to verify correctness up to the
point where the problem is sent to the back-end solver,
and to facilitate the last step, albeit in an unverified way.

We have implemented a \lstinline{solve} command in CvxLean which takes a
an optimization problem \lstinline{prob} in DCP form and carries out the following steps:
\begin{enumerate}
	\item It applies the \lstinline{dcp} procedure to obtain a reduced problem,
  \lstinline{prob.reduced}, and a reduction \lstinline{red : Solution prob.reduced -> Solution prob}.
  \item It carries out the translation to floats, traversing each
  expression and applying the registered translations.
  \item It extracts the numerical data from the problem. At this point, we have scalars, arrays and matrices associated to every type of constraint.
  \item It writes the problem to an external file in the conic benchmark format.\footnote{\url{https://docs.mosek.com/latest/rmosek/cbf-format.html}}
  \item It calls MOSEK and receives a status code in return, together with a solution, if MOSEK succeeds in finding one. The problem status is added to the environment and if it is infeasible or ill-posed, we stop.
  \item Otherwise, the \lstinline{solve} command interprets the solution so that it matches the shape of the variables of \lstinline{prob.reduced}. It also expresses these values as Lean reals,
  resulting in an approximate solution \lstinline{p} to \lstinline{prob.reduced}.
  It declares a corresponding \lstinline{Solution} to \lstinline{prob.reduced},
  using a placeholder for the proofs of feasibility and optimality (since we
  simply trust the solver here).
  \item It then uses the reduction from \lstinline{prob} to \lstinline{prod.reduced},
  again reinterpreted in terms of floats, to compute an approximate solution to
  \lstinline{prob}.
\end{enumerate}
Finally, the results are added to the Lean environment.
In the following example, the command \lstinline{solve so1} results in the creation
of new Lean objects
\lstinline{so1.reduced}, \lstinline{so1.status}, \lstinline{so1.value}, and
\lstinline{so1.solution}. The first of these
represents the conic-form problem that is sent to the back-end solver, while
the remaining three comprise the resulting solution.
\begin{lstlisting}
noncomputable def so1 :=
  optimization (x y : ℝ)
    maximize sqrt (x - y)
    subject to
      c1 : y = 2*x - 3
      c2 : x^2 ≤ 2
      c3 : 0 ≤ x - y

solve so1
#print so1.reduced   -- shows the reduced problem 
#eval so1.status     -- "PRIMAL_AND_DUAL_FEASIBLE"
#eval so1.value      -- 2.101003
#eval so1.solution   -- (-1.414214, -5.828427)
\end{lstlisting}

\section{Related work}
\label{section:related:work}

Our work builds on decades of research on convex optimization \cite{boyd-vandenberghe-2014-convex,rockafellar-1970-convex-analysis,nesterov-2018-convex-optimization,vishnoi-2021-algorithms}, and most directly on the CVX family and disciplined
convex programming \cite{grant-et-al-2006-dcp,grant-boyd-2014-cvx,diamond-boyd-2016-cvxpy,fu-2020-cvxr, udell-et-al-2014-convexjl}. Other popular packages include
Yalmip \cite{lofberg-2004-yalmip}.

Formal methods have been used to solve bounding problems \cite{ratschan-07-safety-verification, gao-et-al-12-delta-complete}, constraint satisfaction problems \cite{franzle-et-al-07-constraint-systems}, and optimization problems \cite{kong-et-al-07-exists-forall}. This literature is too broad to survey here, but \cite{deshmukh-et-al-19-verification-cyber-physical} surveys some of the methods that are used in connection with the verification of cyber-physical systems.
Proof assistants in particular have been used to verify bounds in various ways. Some approaches use certificates from numerical packages; Harrison \cite{harrison-07-sum-of-squares} uses certificates from semidefinite programming in HOL Light, and Magron et al.~\cite{magron:et:al:15} and Martin-Dorel and Roux \cite{dorel-roux-17-valid-sdp} use similar certificates in Coq.
Solovyev and Hales use a combination of symbolic and numeric methods in HOL Light~\cite{solovyev-hales-14-nonlinear-inequalities}. Other approaches have focused on verifying symbolic and numeric algorithms instead. For example, Mu{\~{n}}oz, Narkawicz, and
Dutle \cite{munoz-narkawicz-dutle-18-univariate-pvs} verify a decision procedure for univariate real arithmetic in PVS and Cordwell, Tan, and Platzer \cite{cordwell-tan-platzer-21-univariate} verify another one in Isabelle.
Narkawicz and Mu\~noz \cite{narkawicz-munoz-13-bnb} have devised a verified numeric algorithm to find bounds and global optima. Cohen et al.~\cite{cohen-davy-feron-garoche-17,cohen-feron-garoche-2020-verification-convex} have developed a framework for verifying optimization algorithms using the ANSI/ISO C Specification Language (ACSL) \cite{acsl-2020}.

Although the notion of a convex set has been formalized in a number of theorem provers, we do not know of any full development of convex analysis.
The Isabelle \cite{nipkow-paulson-wenzel-2002} {\sf HOL-Analysis} library
includes properties
of convex sets and functions, including Carath\'eodory's theorem on convex hulls, Radon's theorem, and Helly's theorem, as well as properties of convex sets and functions on normed spaces and Euclidean spaces. A theory of lower semicontinuous functions by Grechuk \cite{grechuk-2011-lower-semicontinuous-afp} in the Archive of Formal Proofs \cite{blanchette-et-al-2015-afp} includes properties of convex functions. Lean's {\sf mathlib} \cite{mathlib-2020} includes a number of fundamental results,
including a formalization of the Riesz extension theorem by Kudryashov and Dupuis and a formalization of Jensen's inequality by Kudryashov. Allamigeon and Katz have formalized a theory of convex polyhedra in Coq with an eye towards applications to linear optimization \cite{allamigeon-katz-2019-formalization-convex-polyhedra}. We do not know of any project that has formalized the notion of a reduction between optimization problems.

\section{Conclusions}
\label{section:conclusions}

We have argued that formal methods can bring additional reliability
and interactive computational support to the practice of convex optimization.
The success of our prototype shows that it is possible to carry out and verify
reductions using a synergistic combination of automation
and user interaction.

The implementation of CvxLean is currently spread between two versions of
Lean \cite{de-moura-et-al-2015-lean,de-moura-ullrich-2021-lean4}.
Lean 3 has a formal library, {\sf mathlib} \cite{mathlib-2020}, which
comprises close to a million lines of code 
and covers substantial portions of algebra, linear algebra, topology, measure theory, and analysis. Lean 4 is a performant programming language as well
as a proof assistant, but its language is not backward compatible with that
of Lean 3.
All of the substantial programming tasks described here have been carried out in Lean 4,
but we rely on a binary translation of the Lean 3 library and some additional results proved there.
This arrangement is not ideal, but a source-level
port of the Lean 3 library is already underway,
and we expect to move the development entirely to Lean 4 in the near future.

There is still a lot to do. We have implemented and verified all the atoms
needed for the examples presented in this paper, but these are still only a
fraction of the atoms that are found in CVXPY.
The DCP transformation currently leaves any side conditions that it cannot prove
for the user to fill in, and special-purpose \emph{tactics},
i.e.~small-scale automation, could help dispel proof obligations like
monotonicity. Textbooks often provide standard methods and tricks for carrying
out reductions (e.g.~\cite[Section 4.1.3]{boyd-vandenberghe-2014-convex}),
and these should also be supported by tactics in CvxLean.
Our project, as well as Lean's library, would benefit from more formal
definitions and theorems in convex analysis and optimization.
We need to implement more efficient means of extracting numeric values
for the back-end solver, and
it would be nice to verify more of the numeric computations and claims.
Finally, and most importantly, we need to work out more examples like the
ones presented
here to ensure that the system is robust and flexible enough to join the
ranks of conventional optimization systems like CVXPY.

{\paragraph{Acknowledgements}
Seulkee Baek did some preliminary experiments on connecting Lean 3 to external optimization solvers.
Mario Carneiro and Gabriel Ebner advised us on how to formalize optimization problems
and on Lean~4 metaprogramming.
Steven Diamond helped us understand the world of convex optimization.
We also had helpful discussions with Geir Dullerud, Paul Jackson, Florian Jarre, John Miller,
Balasubramanian Narasimhan, Ivan Papusha, and Ufuk Topcu.
Steven Diamond, Paul Jackson, and Parth Nobel provided helpful feedback on a draft of this paper.
This work has been partially supported by the Hoskinson Center for Formal
Mathematics at Carnegie Mellon University.
Bentkamp has received funding from a Chinese Academy
of Sciences President’s International Fellowship for Postdoctoral
Researchers (grant No. 2021PT0015).
We thank the anonymous reviewers for their corrections and suggestions.}

\bibliographystyle{splncs04}
\bibliography{bib}

\begin{extended}
\newpage

\section*{Appendix}

In this appendix we provide a more formal description of the proof-producing
DCP reduction algorithm described in Sections~\ref{section:reduction:to:conic:form} and \ref{section:verifying:the:reduction}.

Recall, first, that to each atomic function $f(\bar x)$, we associate the following data:
\begin{itemize}
  \item $\mathsf{expr}_f(\bar{x})$, a Lean expression defining the function
  \item $\mathsf{curv}_f$, its curvature, either $\mathsf{convex}$, $\mathsf{concave}$, or $\mathsf{affine}$
  \item $\mathsf{mono}_{f, i}$, the monotonicity of the function in each of its arguments $i$, either
  $\mathsf{increasing}$, $\mathsf{decreasing}$, $\mathsf{neither}$, or $\mathsf{auxiliary}$ (which is used for parameters to the function that are fixed for the optimization problem)
  \item expressions $\mathsf{constr}_f(\bar{x}, \bar{v})$ and $\mathsf{obj}_f(\bar{x}, \bar{v})$, characterizing the value $f(\bar{x})$ as the minimum value of $\mathsf{obj}_f(\bar{x}, \bar{v})$ under constraints $\mathsf{constr}_f(\bar{x}, \bar{v})$
  \item conditions $\mathsf{bconds}_{f}(\bar{x})$ and $\mathsf{vconds}_{f}(\bar{x})$ on the values where the graph implementation is valid
  \item an expression $\mathsf{sol}_{f,v}(\bar{x})$ for each graph implementation variable $v$,
  providing an optimal solution to the graph implementation for the values where the graph implementation is valid.
\end{itemize}
We also associate with each atom formal proofs of the forward and backward properties indicated below.

Recall also that we use the term \emph{component} to mean
either the objective function or one of the constraints of
a problem.
For each component, we construct a tree where each internal node
corresponds to an atom.
Each node has as many branches as the atom has arguments.
The leaves of the tree are expressions that are affine
with respect to the optimization variables.

We assign an identifier $n$ to each node of the tree
in such a way that the root is identified by the name of the component
and the identifier of the $i$th child node of node $n$
is $n.i$.
We denote the atom at node $n$ by $A(n)$.
For example, in the tree associated with the objective function
\lstinline{-sqrt (x - y)}, the node associated with \lstinline{y}
is identified by $\mathrm{obj}.1.1.2$.

Each node $n$ of a component tree represents the expression
$\mathsf{oexpr}_n(\bar{y})$ in the variables
$\bar{y}$ of the original DCP problem defined as follows.
For each leaf node, $\mathsf{oexpr}_n(\bar{y})$
is the affine expression
associated with that node.
For each inner node,
let $\mathsf{oexpr}_n(\bar{y}) = \mathsf{expr}_{A(n)}(i \mapsto \mathsf{oexpr}_{n.i}(\bar{y}))$
where $\mathsf{expr}_{A(n)}$ is the formal representation of the function associated with the atom.
Here and below, we write $i \mapsto f(i)$ to denote the tuple of expressions whose $i$th entry is $f(i)$.
The tree for each component $\mathrm{comp}$ is constructed so that
the component corresponds to $\mathsf{oexpr}_{\mathrm{comp}}$.
For example, in the tree for
\lstinline{-sqrt (x - y)}, $\mathsf{expr}_{A(\mathrm{obj}.1)} = \text{\lstinline{sqrt}}$
and $\mathsf{oexpr}_{\mathrm{obj}.1} = \text{\lstinline{sqrt (x - y)}}$.

To each node $n$ of a component tree we assign a role $\mathsf{role}_{n}$, either
$\mathsf{convex}$, or $\mathsf{concave}$, or $\mathsf{affine}$.
For the objective function $\mathrm{obj}$, we require
$\mathsf{role}_{\mathrm{obj}} = \mathsf{convex}$,
and for each constraint $\mathrm{constr}_k$, we require
$\mathsf{role}_{\mathrm{constr}_k} = \mathsf{concave}$.
For constraints, the label $\mathsf{concave}$ semantically means
that the constraint describes a convex set; we will
explain the reason for this unfortunate naming below.

The assignment of roles propagates downward through each component's tree,
determined by the monotonicity $\mathsf{mono}_{A,i}$ of each argument $i$ of the atom $A = A(n)$ at each node $n$ and the role $\mathsf{role}_{n.i}$ of the children of $n$:
\begin{itemize}
  \item If $\mathsf{mono}_{A(n),i} = \mathsf{increasing}$, then
    $\mathsf{role}_{n.i} = \mathsf{role}_{n}$.
  \item If $\mathsf{mono}_{A(n),i} = \mathsf{decreasing}$, then
    $\mathsf{role}_{n.i} = - \mathsf{role}_{n}$.
  \item If $\mathsf{mono}_{A(n),i} = \mathsf{neither}$, then
    $\mathsf{role}_{n.i} = \mathsf{affine}$.
  \item If $\mathsf{mono}_{A(n),i} = \mathsf{auxiliary}$, then node $n.i$ is a constant leaf.
\end{itemize}
Here $- \mathsf{convex} = \mathsf{concave}$, $- \mathsf{concave} = \mathsf{convex}$,
and $- \mathsf{affine} = \mathsf{affine}$.
If $\mathsf{curv}_{A(n)} = \mathsf{affine}$,
we can choose $\mathsf{role}_{n}$ arbitrarily to make the tree fulfill the requirements;
otherwise, we must have $\mathsf{role}_{n} = \mathsf{curv}_{A(n)}$.

Some atoms' graph implementations are only valid under certain conditions,
or they fulfill monotonicity or curvature properties only under
certain conditions.
For such atoms, CvxLean introduces
\emph{background conditions} $\mathsf{bconds}_A(\bar{a})$ and
\emph{variable conditions} $\mathsf{vconds}_A(\bar{a})$.
If a tree contains atoms with such conditions,
it is only valid if for all nodes $n$, the statements
$\mathsf{bconds}_{A(n)}(i \mapsto \mathsf{oexpr}_{n.i}(\bar{y}))$ and
$\mathsf{vconds}_{A(n)}(i \mapsto \mathsf{oexpr}_{n.i}(\bar{y}))$
are fulfilled
for all $\bar{y}$. Background conditions and variable conditions
differ in where CvxLean
will search for them.
Variable conditions must be present as constraints of the given optimization problem;
background conditions must be present in the local context of the optimization problem.
Since the local context cannot refer to the optimization variables,
background conditions can only refer to parameters of the atom that are
constant over the domain.

Crucially, the construction of the tree for any constraint is allowed to fail
as long as some other component
requires this constraint as a condition.
For example, the atom $\log(x)$ requires the condition $0 < x$,
and the DCP framework cannot handle strict inequalities.
The graph implementation constraint \lstinline{expCone t 1 x} (representing $\exp(t) \leq x$),
will then be used to justify the constraint $0 < x$, which disappears during the reduction.

Let $\bar{y}$ be the variables of the original problem.
Let $\bar{z}$ be the collection of variables
consisting of one fresh copy for each node's graph implementation variables.
Then the optimization variables of the reduced problem are $\bar{y}, \bar{z}$.
Let $\mathsf{vars}_n(\bar{z})$ be the projection to only those variables in $\bar{z}$
that correspond to node $n$.
To determine the reduced problem, we iterate through each atom tree as
follows, determining a \emph{reduced expression} $\mathsf{rexpr}_n(\bar{y}, \bar{z})$ at each node.
For leaves $n$, let $\mathsf{rexpr}_n(\bar{y}, \bar{z}) = \mathsf{oexpr}_n(\bar{y})$.
For inner nodes $n$, let
$\mathsf{rexpr}_n(\bar{y}, \bar{z})
= \mathsf{obj}_{A(n)}(i \mapsto \mathsf{rexpr}_{n.i}(\bar{y}, \bar{z}), \mathsf{vars}_n(\bar{z}))$.
Notice that even though we write $\mathsf{rexpr}_n(\bar{y}, \bar{z})$,
the only variables among $\bar{z}$ that appear in it are the ones that
have been introduced below node $n$.

The reduced problem's objective function $\mathrm{obj}$ is
$\mathsf{rexpr}_{\mathrm{obj}}(\bar{y}, \bar{z})$.
For each constraint $\mathrm{constr}_k$,
the expression
$\mathsf{rexpr}_{\mathrm{constr}_k}(\bar{y}, \bar{z})$
becomes a constraint of the reduced problem,
unless the constraint is used as a condition of some atom.
Moreover, for each node $n$ occurring in any component's atom tree,
the expression
$\mathsf{constr}_{A(n)}(i \mapsto \mathsf{rexpr}_{n.i}(\bar{y},\bar{z}) , \mathsf{vars}_{n}(\bar{z}) )$
becomes a constraint of the reduced problem.

To establish the equivalence of the original problem and the reduced problem,
we need to store formal proofs of the following with each atom.
\begin{itemize}
  \item Forward properties:
  \begin{itemize}
    \item Solution correctness:\\
    If $\mathsf{vconds}_{A(n)}(\bar{a})$,
    then $\mathsf{obj}_{A(n)}(\bar{a}, \mathsf{sol}_{A(n)}(\bar{a})) = \mathsf{expr}_{A(n)}(\bar{a})$.
    (The equality need not be syntactic here; it suffices that the expressions are provably equal.)

    \item Solution feasibility:
    If $\mathsf{vconds}_{A(n)}(\bar{a})$,
    then $\mathsf{constr}_{A(n)}(\bar{a}, \mathsf{sol}_{A(n)}(\bar{a}))$.

  \end{itemize}
  \item Backward properties:
  \begin{itemize}
    \item Optimality:
    Assume that $\mathsf{constr}_{A(n)}(\bar{a}, \bar{v})$
    for some argument values $\bar{a}$ and some variable values $\bar{v}$.
    Let $\bar{a}'$ be a second tuple of argument values.
    For convex atoms $A(n)$, we need to show: if $\bar{a} \mathrel{\triangle} \bar{a}'$, then $\mathsf{obj}_{A(n)}(\bar{a},\bar{v}) \geq \mathsf{expr}_{A(n)}(\bar{a}')$;
    for concave atoms $A(n)$, we need to to show: if $\bar{a}' \mathrel{\triangle} \bar{a}$, then $\mathsf{obj}_{A(n)}(\bar{a},\bar{v}) \leq \mathsf{expr}_{A(n)}(\bar{a}')$;
    where $\triangle$ denotes ${\geq}$ for increasing arguments,
    ${\leq}$ for decreasing arguments, and
    syntactic equality for neither or auxiliary arguments.
    For affine atoms, we must show both the convex case and the concave case.

    \item Condition elimination:
    If the atom has variable conditions, we must
    prove under the assumptions on $\bar{a}$ and $\bar{a}'$ above that
    $\mathsf{vconds}_{A(n)}(\bar{a}')$ holds.
    This is crucial to eliminate variable conditions that cannot be translated into conic form,
    such as the condition of the logarithm atom.
  \end{itemize}
\end{itemize}
Intuitively, the optimality property combines two different properties of the atom,
namely on the one hand
that the objective function of the graph implementation is bounded by
the atom's function
and on the other hand that the monotonicity properties actually hold.

To establish that the forward and backward properties also hold for predicate atoms (i.e., atoms that return a truth value, like $\le$),
we use Lean's default order \lstinline{false ≤ true}
on elements of the data type of propositions, \lstinline{Prop}.
With this order, registering predicate atoms as $\mathsf{concave}$ amounts to
showing that the predicate describes a convex set.
This unfortunate naming could be avoided by using the order \lstinline{false ≥ true} instead,
but we would like to avoid introducing an additional order on \lstinline{Prop}.

Constructing a strong equivalence requires us to define a forward
map $\varphi$ and a backward map $\psi$.
The definition of the backward map $\psi$ is simple:
since we only add new variables to construct $Q$ from $P$,
we can simply project the domain of $Q$ onto the domain of~$P$.

We now describe the forward map $\varphi$. Let $z_i$ be one
of the new variables of $Q$, say,
introduced as the $j$th variable of the graph implementation of the
atom $A(n)$ and node $n$. Let
\[
  \mathsf{interp}_{z_i}(\bar y) = \mathsf{sol}_{A(n), j}(k \mapsto \mathsf{rexpr}_{n.k}(\bar{y}, \ell \mapsto \mathsf{interp}_{z_\ell}(\bar{y}))),
\]
where, in the inner expression, $z_\ell$ ranges over
the new variables that have been introduced at or below
node $n.k$.
In words, $z_i$ is interpreted as the optimal solution
to the variable introduced by the atom at node $n$,
at the arguments to that atom that are interpreted recursively
in terms of the new variables introduced lower down in the tree.
Assuming the new variables of $Q$ are $z_1, \ldots, z_m$,
write $\mathsf{interp}_{\bar z}(\bar y)$ for the tuple
\[
  \mathsf{interp}_{z_1}(\bar{y}), \dots, \mathsf{interp}_{z_m}(\bar{y}),
\]
and define the forward map $\varphi$ by
\[
  \varphi(\bar y) = (\bar y, \mathsf{interp}_{\bar z}(\bar y)).
\]

To prove, formally, that $(\varphi, \psi)$ is a strong equivalence,
we must show that
for any feasible point $\bar y$ of $P$,
$g (\varphi(\bar y)) \leq f(\bar y)$ and $d (\varphi(\bar y))$.
For the objective function and for the constraints
originating from $P$, we use the
\emph{solution correctness} properties of
the involved atoms.
We can even show the stronger property
that for any feasible point $x$ of $P$,
$g (\varphi(\bar y)) = f (\bar y)$ and
$d_i (\varphi(\bar y)) \Leftrightarrow c_i(\bar y)$
where $d_i$ is a constraint of $Q$ that originates
from a constraint $c_i$ of $P$.

We proceed as follows.
Let $\bar{y}$ be a
feasible point in the domain of the original problem $P$.
We recursively prove
\[
  \mathsf{rexpr}_n(\bar{y},\mathsf{interp}_{\bar z}(\bar y)) = \mathsf{oexpr}_n(\bar{y}).
\]

For leaf nodes, this is true by definition of $\mathsf{rexpr}$.
For inner nodes, we prove the following equalities:
\begin{align*}
\mathsf{rexpr}_n & (\bar{y},\mathsf{interp}_{\bar z}(\bar y)) \\
&= \mathsf{obj}_{A(n)}(i \mapsto \mathsf{rexpr}_{n.i}(\bar{y},\mathsf{interp}_{\bar z}(\bar y)), \mathsf{vars}_{n}(\mathsf{interp}_{\bar z}(\bar y)))
\\&\quad\text{by definition of $\mathsf{rexpr}$}\\
&= \mathsf{obj}_{A(n)}(i \mapsto \mathsf{rexpr}_{n.i}(\bar{y},\mathsf{interp}_{\bar{z}}(\bar{y})), \mathsf{sol}_{A(n)}(i \mapsto \mathsf{rexpr}_{n.i}(\bar{y}, \mathsf{interp}_{\bar{z}}(\bar{y}))))
\\&\quad\text{by definition of $\mathsf{interp}$}\\
&= \mathsf{obj}_{A(n)}(i \mapsto \mathsf{oexpr}_{n.i}(\bar{y}), \mathsf{sol}_{A(n)}(i \mapsto \mathsf{oexpr}_{n.i}(\bar{y})))
\\&\quad\text{by the inductive hypothesis}\\
&= \mathsf{expr}_{A(n)}(i \mapsto \mathsf{oexpr}_{n.i}(\bar{y}))
\\&\quad\text{by solution correctness}\\
&= \mathsf{oexpr}_n(\bar{y})
\\&\quad\text{by definition of $\mathsf{oexpr}$.}
\end{align*}
Above, for solution correctness,
we use that $\bar{y}$ is a
feasible point of $P$ and hence
all atoms' conditions
are fulfilled.
Thus, for the objective function and for the constraints
originating from $P$,
$g (\varphi(\bar y)) = f(\bar y)$ and
$d_i (\varphi(\bar y)) \Leftrightarrow c_i(\bar y)$.

For the constraints introduced due to
constraints of graph implementations of atoms,
we use the \emph{solution feasibility}
property of those atoms.
As above, we assume that $\bar{y}$ is a feasible point
of $P$
and hence
all atoms' conditions
$\mathsf{vconds}_{A(n)}(i \mapsto \mathsf{oexpr}_{n.i}(\bar{y}))$
are fulfilled.
By solution feasibility,
$\mathsf{constr}_{A(n)}(i \mapsto \mathsf{oexpr}_{n.i}(\bar{y}), \mathsf{sol}_{A(n)}(i \mapsto \mathsf{oexpr}_{n.i}(\bar{y})))$.
By the equation derived above, it follows that
$\mathsf{constr}_{A(n)}(i \mapsto \mathsf{rexpr}_{n.i}(\bar{y},\mathsf{interp}_{\bar{z}}(\bar{y})) , \mathsf{sol}_{A(n)}(i \mapsto \mathsf{rexpr}_{n.i}(\bar{y},\mathsf{interp}_{\bar{z}}(\bar{y})) ))$,
which by definition of $\mathsf{interp}$ is equivalent to
\[
  \mathsf{constr}_{A(n)}(i \mapsto \mathsf{rexpr}_{n.i}(\bar{y},\mathsf{interp}_{\bar{z}}(\bar{y})) , \mathsf{interp}_{\bar{z}}(\bar{y}) ).
\]
Thus also the constraints
introduced due to
constraints of graph implementations of atoms
are fulfilled.

The second property we need to show
is that for any feasible point $\bar y, \bar z$ of $Q$,
$f (\bar y) \leq g (\bar y, \bar z)$ and $c (\bar y)$.
For the objective function and
for those constraints of $P$ that were not omitted in $Q$,
we use the \emph{optimality}
property as follows.
We recursively show that
\[\mathsf{rexpr}_{n}(\bar{y},\bar{z})
\mathrel{\square}
\mathsf{oexpr}_{n}(\bar{y})
\]
where $\square$ denotes
$\geq$ if $n$ is convex (or affine in the role of convex),
$\leq$ if $n$ is concave (or affine in the role of concave),
and $=$ if $n$ is affine or a leaf.
At the leaves, the property is obvious by the definition of $\mathsf{rexpr}$.
Affine nodes in the role of affine can always be constructed to be leaves.
At other inner nodes, we apply the \emph{optimality} property.
The conditions of \emph{optimality} are fulfilled
for $\bar{a} = i \mapsto \mathsf{rexpr}_{n.i}(\bar{y},\bar{z})$
and $\bar{a}' = i \mapsto \mathsf{oexpr}_{n.i}(\bar{y})$
by the inductive hypothesis, provided that the tree is valid.
Thus, choosing $\bar{v} = \mathsf{vars}_n(\bar{z})$, we have
\[\mathsf{obj}_{A(n)}(i \mapsto \mathsf{rexpr}_{n.i}(\bar{y},\bar{z}),\mathsf{vars}_n(\bar{z}))
\mathrel{\square} \mathsf{expr}_{A(n)}(i \mapsto \mathsf{oexpr}_{n.i}(\bar{y}))\]
By definition of $\mathsf{rexpr}$ and $\mathsf{oexpr}$, this is equivalent to
$\mathsf{rexpr}_n(\bar{y},\bar{z})
\mathrel{\square} \mathsf{oexpr}_n(\bar{y})$.

For the constraints of $P$ that have been omitted in $Q$,
we use the \emph{condition elimination} property.
To be able to invoke it,
we use the property that we have shown above.
We obtain that
the variable conditions $\mathsf{vconds}_{A(n)}(i \mapsto \mathsf{oexpr}_{n.i}(\bar{y}))$
hold, which are exactly the
constraints of $P$ that have been omitted in $Q$.
\end{extended}

\end{document}